\documentclass[aps,prl,twocolumn,superscriptaddress]{revtex4-1}
\usepackage{braket}
\usepackage{graphicx}
\usepackage{amsmath}    % need for subequations
\usepackage{amsfonts}
\usepackage[many]{tcolorbox}
\usepackage{subfigure}  % use for side-by-side figures
\usepackage{epstopdf}

\begin{document}

\title{Unified quantum theory of energy transfer in nanophotonics:\\ limits on efficiency and energy transfer times}
\title{Fundamental efficiency bound for  \\
 coherent energy transfer  in nanophotonics}
%\title{Quantum-classical bounds of energy transfer efficiency in nanophotonics}
\author{Cristian L. Cortes}
\affiliation{Birck Nanotechnology Center and Purdue Quantum Center, \\ School of Electrical and Computer Engineering,\\ Purdue University, West Lafayette, IN 47906, U.S.A.}
\author{Zubin Jacob}
\affiliation{Birck Nanotechnology Center and Purdue Quantum Center, \\ School of Electrical and Computer Engineering,\\ Purdue University, West Lafayette, IN 47906, U.S.A.}

\begin{abstract}
We derive a unified quantum theory of coherent and incoherent energy transfer between two atoms (donor and acceptor) valid in arbitrary Markovian nanophotonic environments. Our theory predicts a fundamental bound $\eta_{max} = \frac{\gamma_a}{\gamma_d + \gamma_a}$ for energy transfer efficiency arising from the spontaneous emission rates $\gamma_{d}$ and $\gamma_a$ of the donor and acceptor. We propose the control of the acceptor spontaneous emission rate as a new design principle for enhancing energy transfer efficiency. We predict an experiment using mirrors to enhance the efficiency bound by exploiting the dipole orientations of the donor and acceptor. Of fundamental interest, we show that while quantum coherence implies the ultimate efficiency bound has been reached, reaching the ultimate efficiency does not require quantum coherence. Our work paves the way towards nanophotonic analogues of efficiency enhancing environments known in quantum biological systems.

%We present a simple nanophotonic approach to control energy transfer efficiency using the dipole orientations of a donor-acceptor system above a mirror.  %Our theory is exactly soluble and therefore provides an important benchmark for Markovian theories of energy transfer, while also paving the way towards a comprehensive theory of energy transfer in non-Markovian nanophotonic environments.ewr
\end{abstract}

\maketitle
\noindent
Using quantum coherence and correlations as a resource has become a fundamental topic of research in recent years  \cite{adesso2016measures,streltsov2016quantum}. In quantum metrology, quantum correlations are used to go beyond classical measurement limits \cite{lloyd2008enhanced,Beyondclassicallimits,bechera2014overcoming}. In quantum thermodynamics, the use of quantum coherence has been proposed to go beyond the Carnot efficiency limit of classical heat engines \cite{scully2003extracting,dorfman2013photosynthetic,rossnagel2014nanoscale}. And in quantum biology, landmark experiments have shown long-lived coherence on the order of picoseconds suggesting its role in the near-unity energy transfer efficiency of photosynthetic systems \cite{engel2007evidence,collini2010coherently,panitchayangkoon2010long}. The idea of quantum coherence playing a role in photosynthesis is intriguing because it indicates many-body quantum correlations can exist in ambient conditions with the potential for a wide range of technological applications \cite{wallrabe2005imaging,romero2014quantum,bredas2017photovoltaic}. 

%While a complete understanding of photosynthetic energy transfer has not been achieved \cite{kassal2013does}, there has been a lot of progress outlining how the environment can positively influence energy transfer efficiency \cite{kassal2012environment,mohseni2014energy,rebentrost2009environment}.

Energy transfer is typically distinguished as incoherent F\"orster-type resonance energy transfer (FRET), or coherent excitation energy transfer. The two regimes occur in the limits, $J_{dd}/\gamma_{tot} \ll 1$ and $J_{dd}/\gamma_{tot} \gg 1$, involving the ratio of the electronic dipole-dipole coupling $J_{dd}$ to the total linewidth $\gamma_{tot}$ of each molecule. The total linewidth is a measure of the coupling strength to the bath's spin, vibrational or electrodynamic degrees of freedom. In photosynthetic systems, the system-bath coupling is primarily dominated by vibrations. The complex nature of photosynthetic systems results in electronic and vibrational coupling strengths varying greatly between the incoherent and coherent coupling limits. Understanding the role of the environment from the weak-to-intermediate-to-strong coupling regimes has been an important topic of interest required to explain experimental observations \cite{chenu2015coherence}. In this regard, there has been tremendous progress in the development of a wide variety of open quantum system frameworks (modified-Redfield, Hierarchical equations of motion, Polaron-modified master equation) \cite{yang_influence_2002,ishizaki2009unified,wang2015nonequilibrium,beljonne2009beyond,tanimura2014reduced,zimanyi2012theoretical,mccutcheon2011consistent} that operate under a wide range of coupling strengths. While a complete understanding of photosynthetic energy transfer has not been achieved \cite{kassal2013does}, there has been a lot of progress outlining how the environment can positively influence energy transfer efficiency \cite{yang_influence_2002,ishizaki2009unified,wang2015nonequilibrium,beljonne2009beyond,tanimura2014reduced,zimanyi2012theoretical,mccutcheon2011consistent,kassal2012environment,mohseni2014energy,rebentrost2009environment,chin2013role,croce2014natural}. Understanding the fundamental role of quantum coherence remains an open problem in photosynthesis, and it is still not clear whether it does play a role \cite{duan2017nature}. It is possible that other guiding principles give rise to near-unity efficiencies in photosynthesis.

Precise control of resonance energy transfer has also emerged as a fundamental topic of interest in the quantum optics, solid-state and nanophotonics communities. In particular, the description of multi-atom and multi-photon quantum dynamics remains an open challenge in nanophotonics. Theoretically, several authors have proposed the use of plasmonic and nanophotonic systems to enhance energy transfer rates \cite{druger_theory_1987, agarwal_microcavityinduced_1998,bay_atomatom_1997,kobayashi_resonant_1995,govorov_theory_2007,durach_nanoplasmonic_2008,pustovit_resonance_2011}. Several experiments have demonstrated suppression, no-effect, and enhancement of energy transfer rates with plasmonic, optical waveguide, and cavity-based systems \cite{komarala_surface_2008,zhang_enhanced_2007,andrew_forster_2000,blum_nanophotonic_2012,zhao_plasmon-controlled_2012,rabouw_photonic_2014,Tumkur_FRET2015}. Unlike work in the photosynthetic community, most nanophotonic theories of energy transfer have relied on classical electrodynamic descriptions or perturbative approaches based on Fermi's golden rule. While some authors have provided rigorous quantum electrodynamic formulations, the final analytical expressions are typically valid in either the weak or strong coupling regimes \cite{andrews1999resonance,dung2002intermolecular,dung2002resonant}. Moreover, a proper definition of the energy transfer efficiency has been lacking in nanophotonics where most results use F\"orster's perturbative expression. 

In the Letter and supplementary information, we combine ideas from both communities to develop an exactly solvable theory for resonance energy transfer from first-principles. We derive a quantum master equation providing a unified picture of energy transfer dynamics in the coherent and incoherent coupling regimes applicable in arbitrary Markovian nanophotonic environments. We then solve the model exactly to derive a simple analytical expression for the energy transfer efficiency. Our result provides insight into the role of finely-tuned coupling strengths, dephasing rates, and detuning between the donor and acceptor required to achieve near-unity energy transfer efficiencies. The central result of this Letter is the ultimate efficiency of
\begin{equation}
	\eta_{max} = \frac{\gamma_a}{\gamma_d + \gamma_a}.
	\label{ultimateEfficiency}
\end{equation}
This provides a fundamental limit to the energy transfer efficiency between two atoms regardless of coupling strength, quantum coherence and spectral overlap. It also implies the condition $\gamma_a \gg \gamma_d$ is required to achieve near-unity efficiency with the corollary that two identical atoms will have a maximum efficiency of $50 \%$. To the best of our knowledge, this surprisingly simple and intuitive result has not been discussed nor derived in the resonance energy transfer literature. We emphasize this fundamental bound will also apply to quantum transport in the two-chromophore system relevant to many biological systems. 

Interestingly, this bound suggests the acceptor spontaneous emission rate can be used as a new degree of freedom to control energy transfer. To illustrate the interplay of these effects, we predict an experiment to control the efficiency between two atoms above a mirror. We also show that while quantum coherence implies the ultimate efficiency bound has been reached, reaching the ultimate efficiency does not require quantum coherence.  Ultimately, these results will enable the design of nanophotonic systems which can mimic quantum biological environments to enhance energy transfer efficiency.

%Nevertheless, our theory suggests quantum coherence can be used as a resource to achieve \emph{ultra-short} energy transfer times.

%is defined as the incoherent and irreversible transfer of energy from an atom in its excited state (donor) to another atom in its ground state (acceptor). 
%In vacuum, the energy transfer rate has an inverse power law dependence of $r^{-6}$ for two non-overlapping atoms separated by a distance $r$. 
%As the interatomic separation distance decreases, conventional wisdom expects the energy transfer rate to increase and eventually lead to near-unity energy transfer efficiency. 

\paragraph{Perturbative FRET efficiency.} The efficiency of energy transfer is conventionally defined as the ratio of the energy transfer rate $\Gamma_{da}$ to the total dissipation rate of the donor,
\begin{equation}
	\eta_{et} = \frac{\Gamma_{da}}{\Gamma_{da} + \gamma_d}. \nonumber
	\label{PerturbativeEfficiency}
\end{equation}
In free-space, the spontaneous emission rate of the donor is $\gamma_d = d_d^2\omega^3/(3\pi\hbar\epsilon_o c^3)$. The energy transfer rate is $\Gamma_{da} = \frac{2\pi}{\hbar^2}|V_{dd}|^2\mathcal{J}_{da}$ where $\mathcal{J}_{da}$ is the spectral overlap integral of the donor emission and acceptor absorption. The resonant dipole-dipole interaction (RDDI), $V_{dd} = \hbar(-J_{dd}+i\gamma_{dd}/2) = \frac{\omega^2}{\epsilon_o c^2} \mathbf{d}_a\cdot\mathbf{G}(\mathbf{r}_a,\mathbf{r}_d,\omega)\cdot\mathbf{d}_d$, defines the magnitude of the dipole-dipole coupling. The results are written in terms of the dyadic Green function $\mathbf{G}(\mathbf{r}_a,\mathbf{r}_d,\omega)$ containing both near-field Coulombic and far-field radiative contributions. These definitions of the spontaneous emission and energy transfer rates are based on Fermi's Golden rule valid in the incoherent limit. From these relations, we observe that increasing dipole-dipole coupling ($|V_{dd}|\rightarrow \infty$) results in a near-unity energy transfer efficiency, and therefore no fundamental bound exists.

\begin{figure}
	\includegraphics[width=8.5cm]{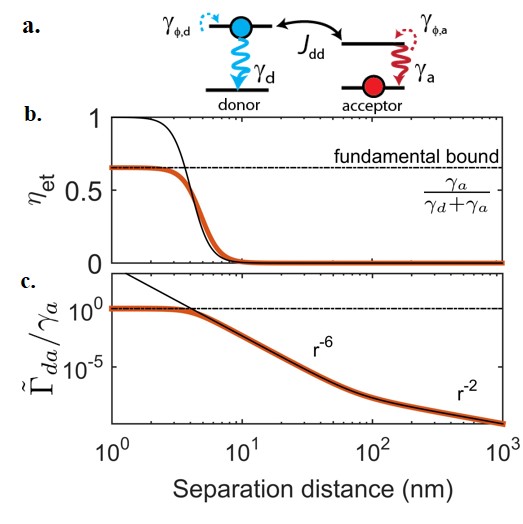}\vspace{-0.2cm}

	\caption{(a) A donor initially in its excited-state will either transfer energy  to an acceptor, or spontaneously emit light with rate $\gamma_{d}$. Once the energy is transferred to the acceptor, the energy can either return to the donor or escape into vacuum with rate $\gamma_{a}$. The energy transfer efficiency is defined as the total probability of an acceptor emitting the initial excitation as opposed to the donor. (b) Using this metric, we find the energy transfer efficiency will have a fundamental bound as the separation distance between two atoms decreases (orange curve), in stark contrast to the conventional definition for the FRET efficiency (black curve). (c) The result can also be understood in terms of the renormalized transfer rate $\tilde{\Gamma}_{da}$ (orange curve) having a fundamental bound as compared to the energy transfer rate $\Gamma_{da}$. We take $\gamma_a = 2\gamma_d$ giving an ultimate efficiency of $\eta_{max} = 2/3$.% Cooperative decay $\gamma_{dd}$ is included in (c).
	}
	\vspace{-0.5cm}
\end{figure}

\paragraph{Non-Perturbative energy transfer efficiency.} In this Letter, we follow the extensive work of photosynthetic excitation energy transfer \cite{olaya2008efficiency,mohseni2008environment} and use the following definition for the energy transfer efficiency,  
\begin{equation}
	{\eta}_{et} = \gamma_a \int_0^\infty\!\!\!\! \rho_{aa}(t) dt, \nonumber
\end{equation}
valid for non-stationary processes such as when the donor is initially in its excited state. This result is general enough to work in the weak and strong coupling regimes between two atoms. Here, the energy transfer efficiency is proportional to the time-integrated luminescence originating from the acceptor. $\rho_{aa}(t)$ is the time-dependent density matrix population of the acceptor in the excited-state. For many applications, this is a much more useful and intuitive definition for the energy transfer efficiency.

In the supplementary information, we derive the RDDI  master equation for two non-identical atoms of the form, $\frac{\partial}{\partial t}\rho = i[\rho,H_{coh}]  + \mathcal{L}[\rho]$, from first principles. The first term involves the coherent dynamics due to dipole-dipole coupling $J_{dd}$. The second term is a Lindblad superoperator describing the incoherent dynamics due to spontaneous emission and pure dephasing of the donor and acceptor respectively. For rest of the Letter, we will ignore non-local cooperative decay $\gamma_{dd}$ typically associated with superradiant and subradiant effects. We will explore these effects in a future paper. Our results are general enough to work in any Markovian bath with a correlation time $\tau_c$ that is much smaller than the relaxation times of the atoms, $\tau_c^{-1} \gg \gamma_d,\gamma_a,\Gamma_{da}$. This extends the range of applicability of this approach beyond the vacuum case, allowing the consideration of more complicated nanophotonic environments. Using the RDDI master equation, a central result of this Letter is the exact analytical expression of the energy transfer efficiency valid in the coherent and incoherent coupling regimes,
\begin{equation}
 	\eta_{et} = \frac{\tilde{\Gamma}_{da}}{\tilde{\Gamma}_{da} + \gamma_d}
\end{equation}
where we define the renormalized energy transfer rate,
\begin{equation}
	\tilde{\Gamma}_{da} = \frac{\gamma_a\Gamma_{da}}{\gamma_a + \Gamma_{da}}.
	\label{TIrate}
\end{equation}
Surprisingly, we recover the same functional form of F\"orster's perturbative energy transfer rate, $\Gamma_{da} = \frac{2\pi}{\hbar^2}|V_{dd}|^2\mathcal{J}_{da}$, however, the master equation approach allows for an exact solution of the spectral overlap integral,
\begin{equation}
 	\mathcal{J}_{da} = \frac{(\gamma_d + \gamma_{\phi,d} + \gamma_a + \gamma_{\phi,a})/(2\pi)}{(\tilde{\omega}_d -\tilde{\omega}_a)^2 + (\gamma_d + \gamma_{\phi,d} + \gamma_a + \gamma_{\phi,a})^2/4}.
 	\label{OverlapIntegral}
\end{equation} 
The overlap integral $\mathcal{J}_{da}$ is equal to the integral of two Lorentzians with resonant frequencies $\tilde{\omega}_d=\omega_d + \delta\omega_d$, $\tilde{\omega}_a=\omega_a+ \delta\omega_a$ and linewidths $\gamma_d + \gamma_{\phi,d}$, $\gamma_a + \gamma_{\phi,a}$ respectively. Here, we introduce $\gamma_{\phi,i}$ as the phenomenological dephasing rate for each atom accounting for fluctuations in the transition frequency. The dephasing rate contributes to an observable linewidth broadening occurring at finite temperatures where $\gamma_{\phi,i} \gg \gamma_i$.

While the functional form for the energy transfer rate $\Gamma_{da}$ is similar to conventional FRET theory, this approach goes beyond the perturbative result by taking into account the modification of the resonant frequency and linewidth of each atom, $\delta\omega_i = -\frac{\omega^2 }{\hbar\epsilon_o c^2}\mathbf{d}_i\cdot\text{Re}\,\mathbf{G}(\mathbf{r}_i,\mathbf{r}_i,\omega)\cdot\mathbf{d}_i$ and $ \gamma_i = \frac{2\omega^2 }{\hbar\epsilon_o c^2}\mathbf{d}_i\cdot\text{Im}\,\mathbf{G}(\mathbf{r}_i,\mathbf{r}_i,\omega)\cdot\mathbf{d}_i$, resulting in non-perturbative emission and absorption spectra for the donor and acceptor respectively. In general, the dyadic Green function consists of vacuum and scattered contributions, reinforcing the applicability of this approach to more complicated nanophotonic environments. 

\begin{figure}[t!]
	\includegraphics[width=9cm]{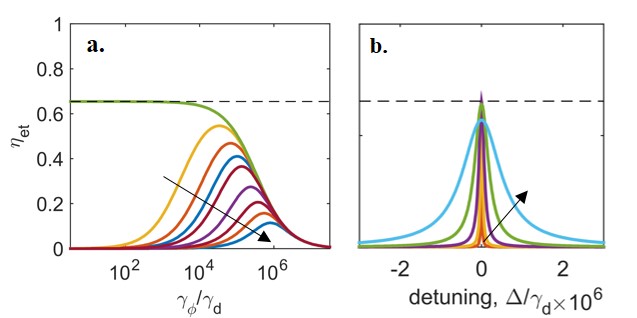}
	\caption{Energy transfer efficiency as function of (a) dephasing rate $\gamma_{\phi}$ and (b) atom-atom detuning $\Delta = \tilde{\omega}_d - \tilde{\omega}_a$. Note the energy transfer efficiency always remains below the fundamental bound regardless of coupling strengths, spontaneous emission, dephasing or detuning. This bound may be reached asymptotically for the case of two atoms with zero detuning in the limit of small dephasing (green curve left). Black arrow denotes (a) increased detuning and (b) increased dephasing.}
\end{figure}

\paragraph{Maximum energy transfer efficiency.} The renormalized energy transfer rate (\ref{TIrate}) arises from the exact non-stationary solution for two non-identical atoms. The perturbative expression for the FRET efficiency  can be recovered when $\Gamma_{da} \ll \gamma_a$. This condition suggests F\"orster's result is only valid when the acceptor has a fast enough dissipation rate to ensure irreversible energy transfer. In realistic systems, the finite dissipation rate of the acceptor will result in a bottleneck effect. Energy cannot be transferred efficiently at a rate faster than the dissipation rate of the acceptor. In the limit of large dipole-dipole coupling, $|V_{dd}|\rightarrow \infty$, the renormalized transfer rate is bounded, $\tilde{\Gamma}_{da} \rightarrow \gamma_a$. The ultimate bound (\ref{ultimateEfficiency}) for the energy transfer efficiency immediately follows. 

The results for the non-perturbative efficiency $\eta_{et}$ and the renormalized transfer rate $\bar{\Gamma}_{da}$ are shown in Fig. (1) for two atoms in vacuum as a function of separation distance. The renormalized transfer rate $\bar{\Gamma}_{da}$ has a $r^{-6}$ inverse power law dependence until it reaches the bottleneck limit of $\gamma_a$, at which point the energy transfer efficiency reaches the fundamental bound. For comparison, we plot the energy transfer efficiency as would be predicted through F\"orster's expression (black line).

In figure 2, we provide numerical evidence of the robustness of this bound to atom-atom detuning $\Delta = \tilde{\omega}_d - \tilde{\omega}_a$ as well as dephasing. It is shown that the fundamental efficiency bound can only be approached in the limit of zero detuning, $\Delta\rightarrow 0$. For large detuning, the energy transfer rate will decrease due poor spectral overlap in the absence of dephasing. As dephasing is increased the energy transfer efficiency reaches a maximum (see Fig 2-a) when the following condition is satisfied
\begin{equation}
	(\tilde{\omega}_d -\tilde{\omega}_a)^2 = (\gamma_d  + \gamma_a + 2\gamma_{\phi})^2/4.
\end{equation}
Here, we have assumed equal dephasing for both atoms, $\gamma_\phi = \gamma_{\phi,d} = \gamma_{\phi,a}$. Condition (5) corresponds to the optimal emission-absorption spectral overlap. The use of dephasing to enhance efficiency is often referred to as environment assisted quantum transport (ENAQT). 
%In the limit $\gamma_\phi \rightarrow \infty$, the quantum Zeno effect results in zero energy transfer rate with negligible efficiency. 

\paragraph{Role of quantum coherence.} In general, quantum coherence is achieved in the strong coupling regime, $|V_{dd}| \gg \gamma_d,\gamma_a,\gamma_\phi$. The strong coupling condition coincides with the condition required to achieve fundamental bound (1), $|V_{dd}|\rightarrow \infty$, therefore any system with strong coupling and quantum coherence will operate at an efficiency equal to the fundamental bound (1). However, we emphasize the opposite is not true: operating near the fundamental bound does not imply the system has quantum coherence. To demonstrate this effect, we show the population dynamics and efficiency of two distinct systems. We utilize concurrence as a measure of entanglement and quantum coherence in the two-particle system. In fig 3-a, the system consists of a perfectly tuned donor-acceptor pair, $\Delta = 0$, with zero dephasing. This system achieves the ultimate efficiency of $\eta_{max} = 2/3$. The time-dependent concurrence (bottom plot) clearly shows quantum coherence is present in this system. In figure 3-b, we have two detuned atoms $\Delta/(2\pi) = 10$ THz with large dephasing $\gamma_\phi/(2\pi) = 4$ THz close to the necessary condition (5) for optimal spectral overlap. Interestingly, the second system exhibits irreversible energy transfer with negligible concurrence and therefore lacks quantum coherence but nevertheless reaches an efficiency that lies within 1 percent of the fundamental bound. The clear advantage of quantum coherence is that it reaches $\eta_{max}$ for longer distances, $r = 45$ nm, while the detuned system requires a separation distance of $r = 4.5$ nm.  

\begin{figure}[t!]
	\includegraphics[width=8.5cm]{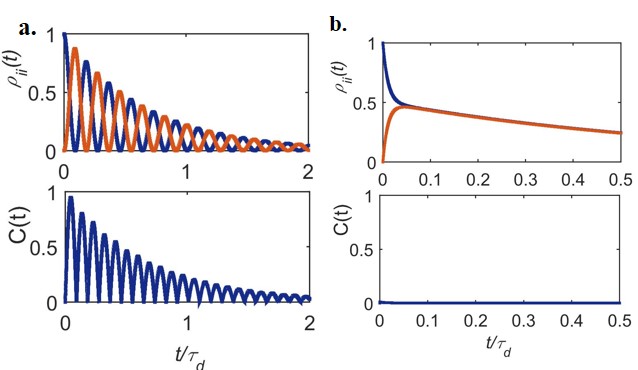}
	\caption{Population dynamics of donor (blue) and acceptor (orange) as well as concurrence (bottom) used as a measure of quantum coherence. (a) Quantum coherent energy transfer between two atoms ($r = 45$ nm) operating at the ultimate efficiency $\eta_{max} = 2/3$. (b) Irreversible energy transfer between two atoms ($r=4.5$ nm) operating within 1 percent of the ultimate efficiency exhibiting negligible quantum coherence. }
\end{figure}

\paragraph{Nanophotonic control of energy transfer efficiency.} The fundamental bound (1) suggests a new design strategy for increasing the energy transfer efficiency based on control of donor and acceptor spontaneous emission rates. In figure 4, we present a canonical example illustrating how a nanophotonic environment can positively influence the energy transfer efficiency between two atoms using a non-resonant mirror eliminating the need for high-Q cavities. The basic idea is to use an orientation-dependent Purcell effect close to the mirror, understood through an image dipole model (inset). A parallel dipole close to a mirror will form an image dipole with the opposite orientation suppressing spontaneous emission, while a perpendicular dipole close to a mirror will form a collinear image dipole enhancing spontaneous emission. This suggests an ideal configuration where the donor is parallel and acceptor is perpendicular to the mirror surface (orange curve). The mirror-enhanced efficiency bound is reached at approximately 10 nm from the mirror. Note that this configuration is typically forbidden in free-space, but becomes possible due to image dipole formation.

\begin{figure}
	\includegraphics[width=9cm]{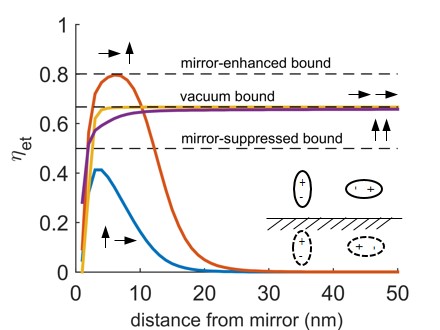}
	\caption{Nanophotonic control of energy transfer between two atoms above a silver mirror. Here, we provide an example of how the environment can positively or negatively influence the energy transfer efficiency based primarily on the transition dipole moment orientation. We consider two atoms with spontaneous emission rates $\gamma_{a} = 2\gamma_{d}$ corresponding to a vacuum bound of $\eta_{max} = 2/3$. To overcome the vacuum bound, we propose using the orientation dipole moments of each atom relative to the mirror to control spontaneous emission rates. The ideal configuration corresponds to a donor parallel to a mirror and an acceptor perpendicular to a mirror, as it achieves the condition $\gamma_a \gg \gamma_d$ around 10 nm from the mirror. In this scenario, the environment modifies the fundamental bound of the energy transfer efficiency resulting in an overall enhancement. The opposite configuration (blue) will decrease the fundamental bound suppressing the overall energy transfer efficiency. Results are calculated with the full dyadic Green function for two atoms $r=10$ nm apart. }
\end{figure}

\paragraph{Conclusion.} To conclude, we derive a fundamental efficiency bound for resonance energy transfer between two atoms in the limit of large dipole-dipole coupling. We use the bound to derive design principles for controlling resonance energy transfer in nanophotonics and present an exactly solvable canonical example to illustrate the interplay of these effects. Our results will be critical in understanding the role of the environment in resonance energy transfer using nanophotonic and metamaterial approaches \cite{cortes2017super}. Future work will focus on developing a rigorous non-Markovian theory of energy transfer with a wider range of applicability to electrodynamic reservoirs.

\bibliography{Theory3}

\end{document}